\documentclass[aps,prb,twocolumn,groupedaddress,showpacs]{revtex4}

\usepackage{graphicx}
\usepackage{dcolumn}
\usepackage{bm}

\begin{document}

\title{On the Imbalanced d-wave Superfluids within the Spin Polarized Extended Hubbard Model: Weak Coupling Limit}

\author{Agnieszka Kujawa-Cichy}
\email{agnieszkakujawa2311@gmail.com}
\affiliation{Solid State Theory Division, Faculty of Physics, Adam Mickiewicz University, Umultowska 85,
61-614 Pozna\'n, Poland}
\date{\today}

\begin{abstract}
We investigate the superfluid properties of d-wave pairing symmetry within the Extended Hubbard Model (EHM) in a magnetic field. We analyze the temperature and magnetic field dependencies of the order parameter. We find that in the two-dimensional case, the spatially homogeneous spin polarized superfluidity ($SC(P\neq0)$) is stable in the weak coupling limit, at $T=0$, as opposed to the s-wave pairing symmetry case in 2D. We construct the ground state phase diagrams both for fixed chemical potential ($\mu$) and electron concentration ($n$). Furthermore, we obtain the temperature vs. magnetic field and temperature vs. spin polarization phase diagrams.
\end{abstract}

\maketitle

\section{Introduction}

In this paper, we briefly discuss the superfluid properties of the Extended Hubbard Model (EHM) with spin independent hopping integrals ($t^{\uparrow}=t^{\downarrow}$), in a Zeeman magnetic field ($h$). We take into account only the pure d-wave pairing symmetry case. Our motivation to study this kind of pairing symmetry is not only the interest in high-temperature superconductivity, but also the possibility of existence of new phases with non-trivial Cooper pairing mechanism in imbalanced Fermi gases. 

There has been much experimental \cite{ketterle2} and theoretical \cite{sarma, Sheehy, Wilczek2, Iskin, parish, kujawa2, ACichy} work on the possibilty of existence of the spatially homogeneous spin-polarized superfluidity (Sarma phase or breached pairing state (BP)) with one or two Fermi surfaces (BP-I or BP-II, respectively) and a gapless spectrum for the majority spin species. The coexistence of the superfluid and the normal component in the isotropic state is characteristic for the BP phase. According to many investigations \cite{parish, ACichy}, at $T=0$, the Sarma phase (or BP-II state) in the weak coupling limit is unstable for the s-wave pairing symmetry case.

The model Hamiltonian is the Extended Hubbard model (EHM) \cite{MicnasModern} in a magnetic field with spin independent hopping integrals:
\begin{eqnarray}
H&=&\sum_{i,j,\sigma}(t_{ij}-\mu\delta_{ij})c_{i\sigma}^{\dag}c_{j\sigma}+U\sum_{i}n_{i\uparrow}n_{i\downarrow}+\nonumber\\
&+&\frac{1}{2}\sum_{i,j,\sigma,\sigma'}W_{ij}n_{i\sigma}n_{j\sigma'}
-h\sum_{i}(n_{i\uparrow}-n_{i\downarrow}),
 \end{eqnarray}
where: $t_{ij}$ -- nearest-neighbor hopping; $\sigma=\uparrow ,\downarrow$ -- spin index, $n_{i\sigma}=c_{i\sigma}^{\dag}c_{i\sigma}$ -- particle number operator, $U$ -- on-site interaction, $W_{ij}$ -- intersite interaction, $h$ -- Zeeman magnetic field, $\mu$ -- chemical potential. 
The gap parameter is defined by: $\Delta_{\vec{k}} =\frac{1}{N}\sum_{\vec{q}} V_{\vec{k}\vec{q}}^{s} \langle c_{-\vec{q} \downarrow} c_{\vec{q} \uparrow} \rangle$, where: $V_{\vec{k}\vec{q}}^{s}=-U-W\gamma_{\vec{k}-\vec{q}}$.  

Applying the broken symmetry Hartree-Fock approximation, we obtain the grand canonical potential $\Omega$ and the free energy $F$. Using the free energy expression, one gets the equations for the gap: $\Delta_{\vec{k}}=\frac{1}{N}\sum_{\vec{q}}V_{\vec{k}\vec{q}}^s \frac{\Delta_{\vec{q}}}{2\omega_{\vec{q}}}(1-f(E_{\vec{q}\uparrow})-f(E_{\vec{q}\downarrow}))$, particle number (which determines $\mu$): $n=n_{\uparrow}+n_{\downarrow}$, $n_{\sigma}=\frac{1}{N} \sum_{\vec{k}} \langle c_{\vec{k} \sigma}^{\dag} c_{\vec{k} \sigma} \rangle=\frac{1}{N}\sum_{\vec{k}}(|u_{\vec{k}}|^2 f(E_{\vec{k}\sigma})+|v_{\vec{k}}|^2 f(-E_{\vec{k}-\sigma}))$, Fock parameter: $p=\frac{p_{\uparrow}+p_{\downarrow}}{2}$, $p_{\sigma}=\frac{1}{N} \sum_{\vec{k}}\gamma_{\vec{k}} \langle c_{\vec{k} \sigma}^{\dag} c_{\vec{k} \sigma} \rangle=\frac{1}{N}\sum_{\vec{k}} \gamma_{\vec{k}}(|u_{\vec{k}}|^2 f(E_{\vec{k}\sigma})+|v_{\vec{k}}|^2 f(-E_{\vec{k}-\sigma}))$ and spin magnetization: $M=n_{\uparrow}-n_{\downarrow}$, where: $f(E_{\vec{k}\sigma})=1/(\exp(\beta E_{\vec{k}\sigma})+1)$, $\beta=1/k_B T$, $E_{\vec{k}\downarrow, \uparrow}= \pm \frac{UM}{2}\pm \frac{1}{2}W(p_{\uparrow}-p_{\downarrow})\frac{\gamma_{\vec{k}}}{\gamma_0} \pm h+\omega_{\vec{k}}$, $\omega_{\vec{k}}=\sqrt{((\epsilon_{\vec{k}}-\bar{\mu})^2+|\Delta_{\vec{k}}|^2}$, $|v_{\vec{k}}|^2=\frac{1}{2} \Big(1-\frac{\epsilon_{\vec{k}}-\bar{\mu}}{\omega_{\vec{k}}}\Big)$, $|u_{\vec{k}}|^2=1-|v_{\vec{k}}|^2$, $\epsilon_{\vec{k} }=-2t\Theta_{\vec{k}}$, $\gamma_{\vec{k}}=2\Theta_{\vec{k}}$, $\Theta_{\vec{k}}=\sum_{l=1}^{d} \cos(k_l a_l)$ ($d=2$ for two-dimensional lattice), $a_l=1$ in further considerations, $\bar{\mu}=\mu-n(\frac{U}{2}+W\gamma_0)$.

We take into account only the pure d-wave pairing symmetry case. Then, $W<0$, $U=0$ and the equation for the order parameter takes the form:
\begin{equation}
 \frac{4}{|W|}=\frac{1}{N} \sum_{\vec{k}} \frac{\eta_{\vec{k}}^2}{2\omega_{\vec{k}}}(1-f(E_{\vec{k}\uparrow})-f(E_{\vec{k}\downarrow})),
\end{equation}
where: $\eta_{\vec{k}}=2(\cos k_x-\cos k_y)$.  

We also calculate the superfluid density $\rho_s(T)$, which takes the form:
\begin{equation}
\rho_s(T)=-\frac{t}{N} \sum_{\vec{k}} \Bigg(\frac{\epsilon_{\vec{k}}-\bar{\mu}}{2\omega_{\vec{k}}} \cos k_x X_{\vec{k}}+t\sin^2k_x Y_{\vec{k}}\Bigg),
\end{equation}
where:
\begin{equation}
 X_{\vec{k}}=\frac{\sinh(\beta \omega_{\vec{k}})}{\cosh(\beta(h+\frac{UM}{2}))+\cosh(\beta \omega_{\vec{k}})},
\end{equation}
\begin{equation}
 Y_{\vec{k}}=\beta \frac{\cosh(\beta(h+\frac{UM}{2}))\cosh(\beta \omega_{\vec{k}})+1}{\big(\cosh(\beta(h+\frac{UM}{2}))+\cosh(\beta \omega_{\vec{k}})\big)^2}.
\end{equation}
The Kosterlitz-Thouless temperature ($T_c^{KT}$) is determined in $d=2$ from the universal relation:
\begin{equation}
\label{KT}
 k_B T_c^{KT}=\frac{\pi}{2} \rho_s (T_c^{KT}).
\end{equation}

\section{Numerical Results}
\begin{figure}[h!]
\includegraphics[width=0.3\textwidth,angle=270]{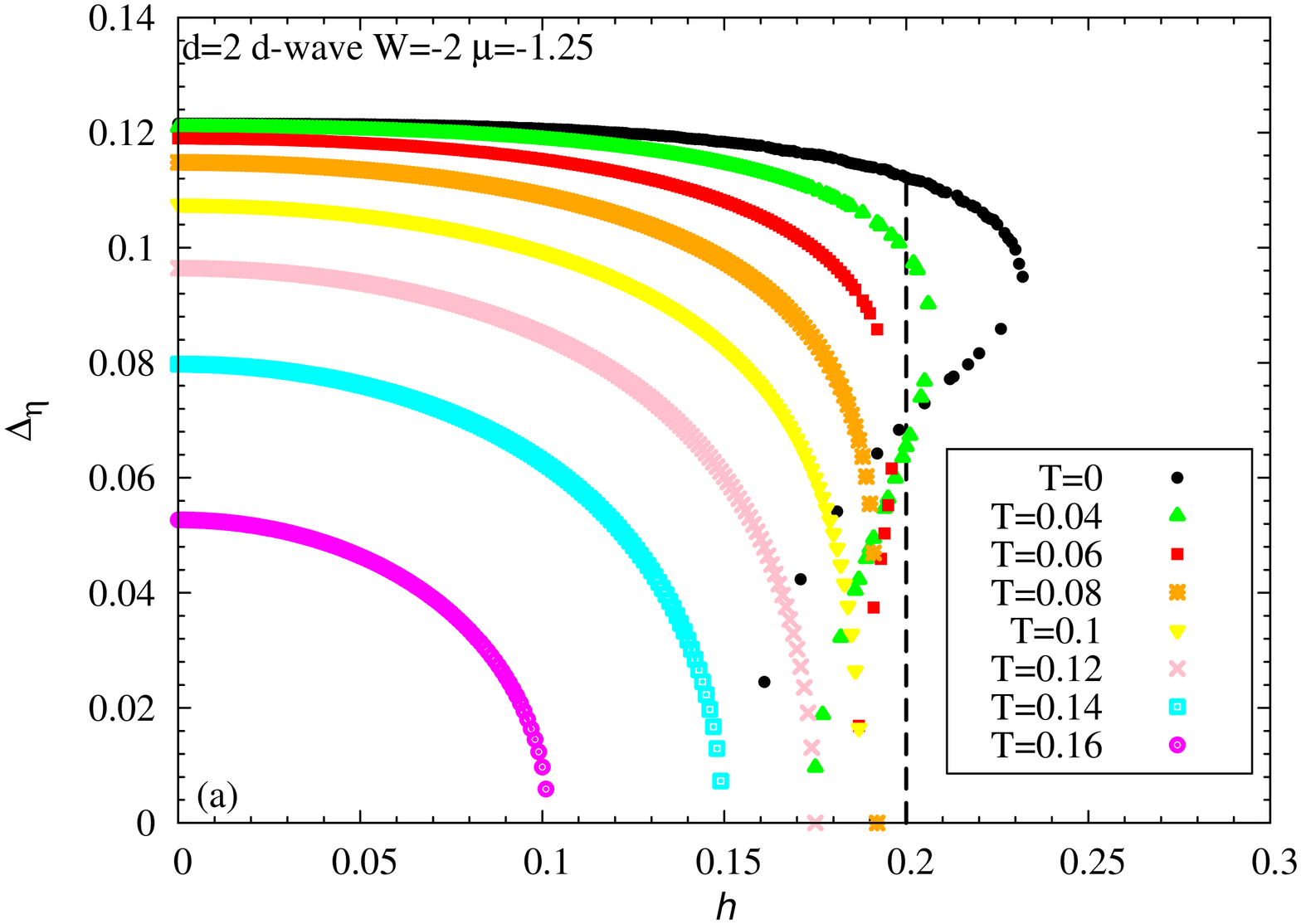}
\includegraphics[width=0.3\textwidth,angle=270]{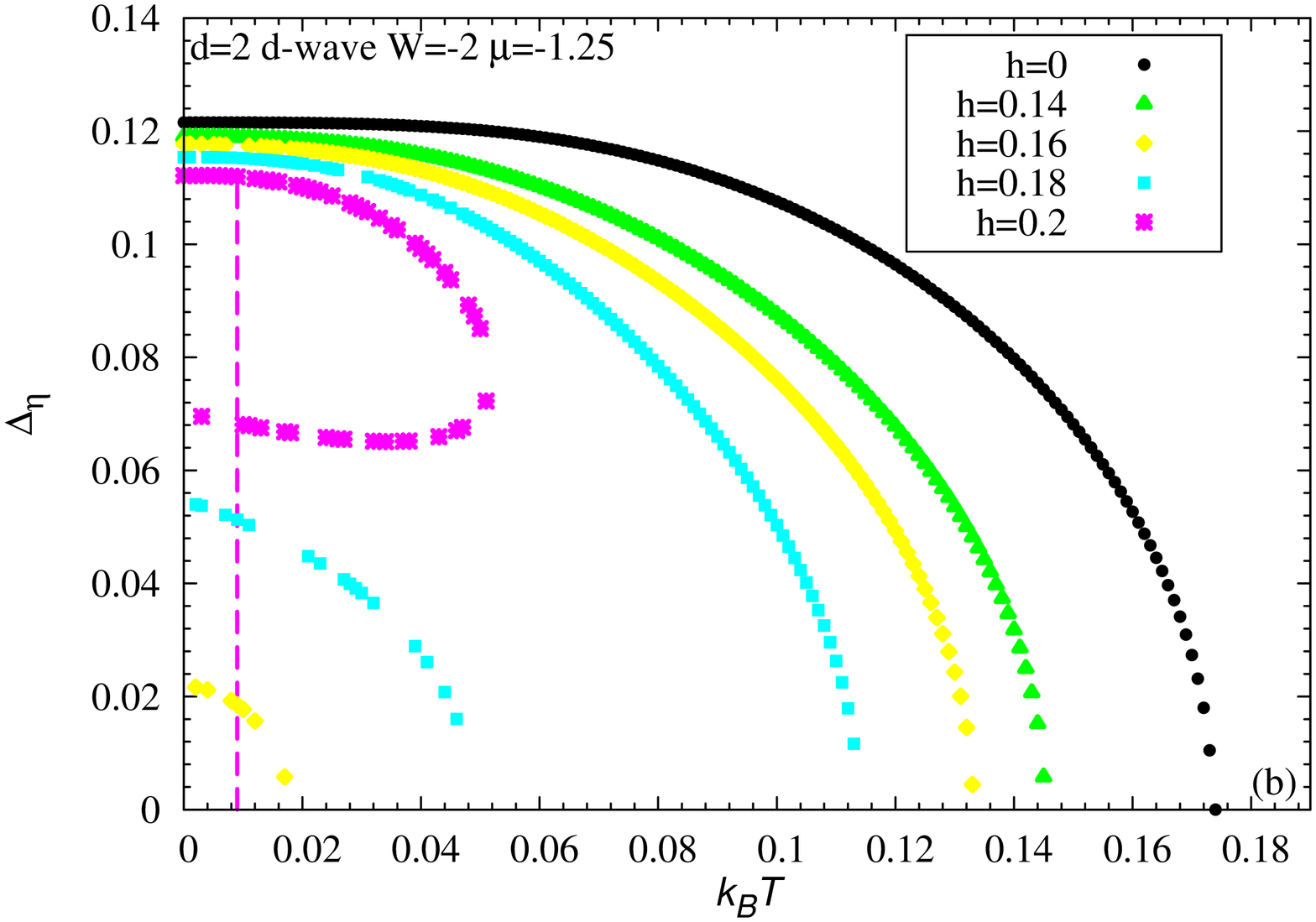}
\caption{\label{del_eta} Dependence of $\Delta_{\eta}$ on the magnetic field (a) and temperature (b), $W=-2$, for a fixed $\mu=-1.25$. In Fig (b), for $h=0.16$ and $h=0.18$ the lower branches are unstable. For $T=0$ (a) and $h=0.2$ (b) the vertical dashed lines denote the first order phase transition from the magnetized superconductor state (SC$(P\neq0)$) to the normal state (NO).}
\label{fig1}
\end{figure}

One of the most important quantities related to superconductivity is the gap
parameter. As is well known, the BCS theory predicts the existence of an
isotropic order parameter, which vanishes at the temperature of the
superconductor-normal phase transition. However, intensive studies of the gap
parameter for high-T$_c$ superconductors indicate significant differences
with respect to the predictions of the BCS theory. Most of the measurements show
that its value in the ground state is much larger than the value of $\Delta$ in
conventional superconductors \cite{Maple}. The symmetry of the energy gap can be determined
from measurements of the changes in its magnitude for different momentum
directions $|\Delta_{\vec{k}}|$. Most studies indicate the $d_{x^2-y^2}$ pairing
symmetry (with the energy gap $\Delta_{\vec{k}}=\Delta_{\eta} (\cos (k_x)-\cos
(k_y))$) \cite{Shen}. 

The d-wave pairing symmetry is also very interesting from the point of view of
the BP state (or Sarma phase) in imbalanced ultracold Fermi
gases.

We start from the analysis of the influence of the magnetic field on the order parameter characteristics.

If $h=0$, in the case of the d-wave pairing symmetry, the order parameter vanishes for some values of the wave vector $\vec{k}$, i.e. along the lines $|k_x|=|k_y|$ (in four nodal points on the Fermi surface). Disappearance of the gap on the Fermi surface leads to the existence of zero energy quasiparticles. It is justified to believe that the Sarma-type phase will be stable at $h\neq 0$, in the weak coupling limit, as opposed to the s-wave pairing symmetry case in 2D.

Fig. \ref{del_eta} shows the dependence of the order parameter amplitude ($\Delta_{\eta}$) on the magnetic field (a) and temperature (b), for $W=-2$, $\mu =-1.25$. As one can see in Fig. \ref{del_eta}(a), there are two different branches of the solutions for $\Delta_{\eta} \neq 0$, at $T=0$, as in the s-wave pairing symmetry case \cite{Kujawa}. However, as opposed to the isotropic order parameter case, the upper branch of the solutions of $\Delta_{\eta}$ is dependent on the magnetic field in the ground state. Therefore, a finite polarization ($P=M/n$) occurs in the system, for arbitrarily small value of the magnetic field, even at $T=0$. This is explained by the creation of polarized quasiparticle excitations in the nodal points of the gap \cite{Tempere'', Yang}. 
Moreover, this branch is stable up to $h\approx 0.2$. At this point, the first order phase transition from the polarized superconducting to the normal state occurs. On the other hand, the lower branch, which also depends on $h$, is unstable at $T=0$. Obviously, the polarization increases with increasing temperature. Thus, the range of occurrence of the Sarma-type phase (the superconducting state with $P\neq 0$) increases.

As mentioned above, the d-wave pairing symmetry at $h=0$ is gapless in four nodal points on the Fermi surface. In the weak coupling limit, at $h=0$, the nodal points are located at ($\pm \frac{\pi}{2}$, $\pm \frac{\pi}{2}$). However, the gap ($E_g$) in the density of states is fixed by the location of the logarithmic singularities. The value of $E_g$ is determined
by the maximum value of the energy gap: $E_g=2\Delta_{max}$, where $\Delta_{max}=4\Delta_{\eta}$.

The influence of the Zeeman magnetic field on the density of states is significant. If $h\neq 0$, the densities of states are
different for the particles with spin ``up'' and spin ``down''. The gap appears in the density of states for the minority spin species and equals $2h$ \cite{kujawa3}. The occurrence of the gap in the density of states is
caused by the existence of some minimum non-zero quasiparticle energy. 
 
\begin{figure}[h!]
\includegraphics[width=0.3\textwidth,angle=270]{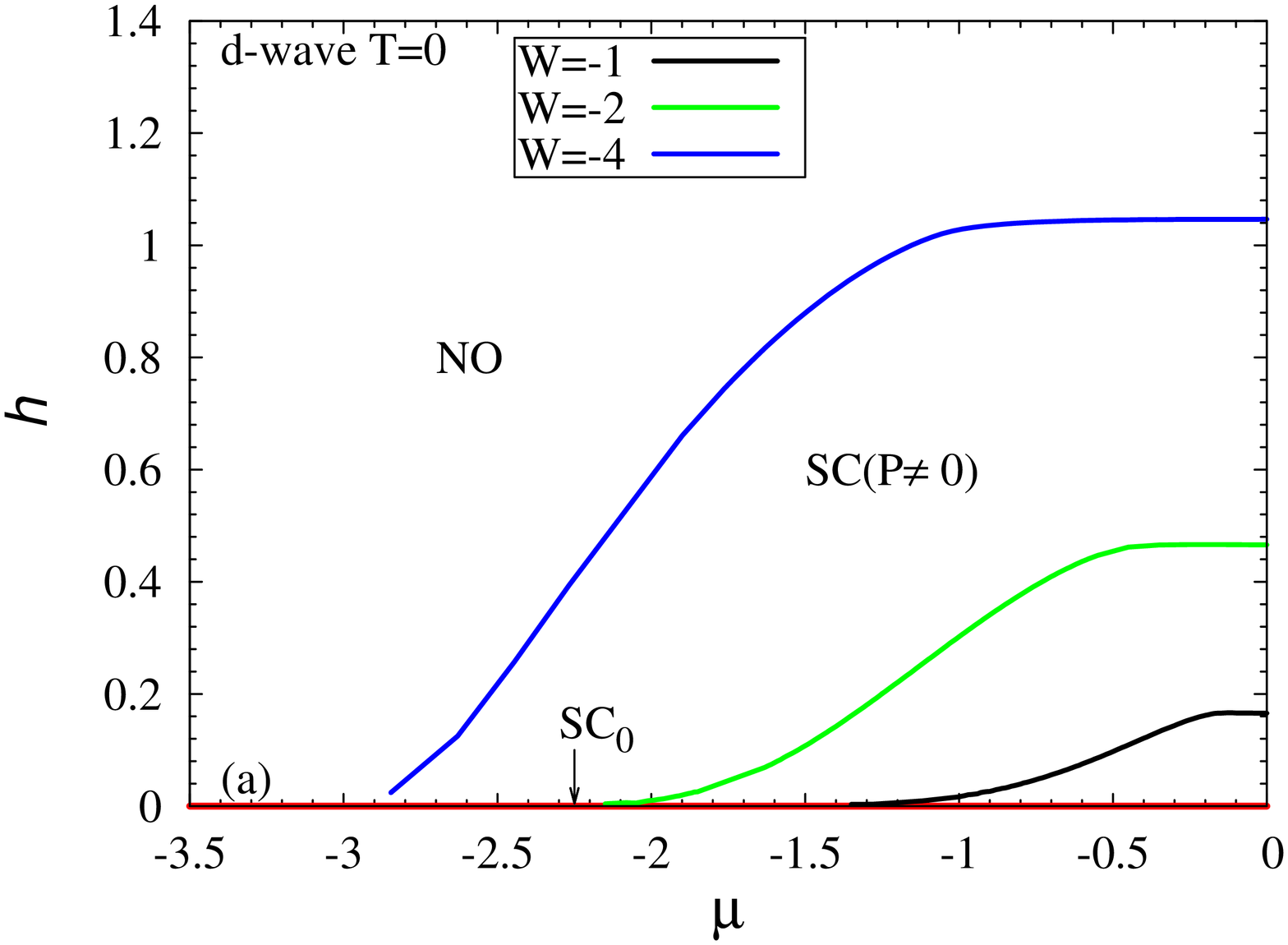}
\includegraphics[width=0.3\textwidth,angle=270]{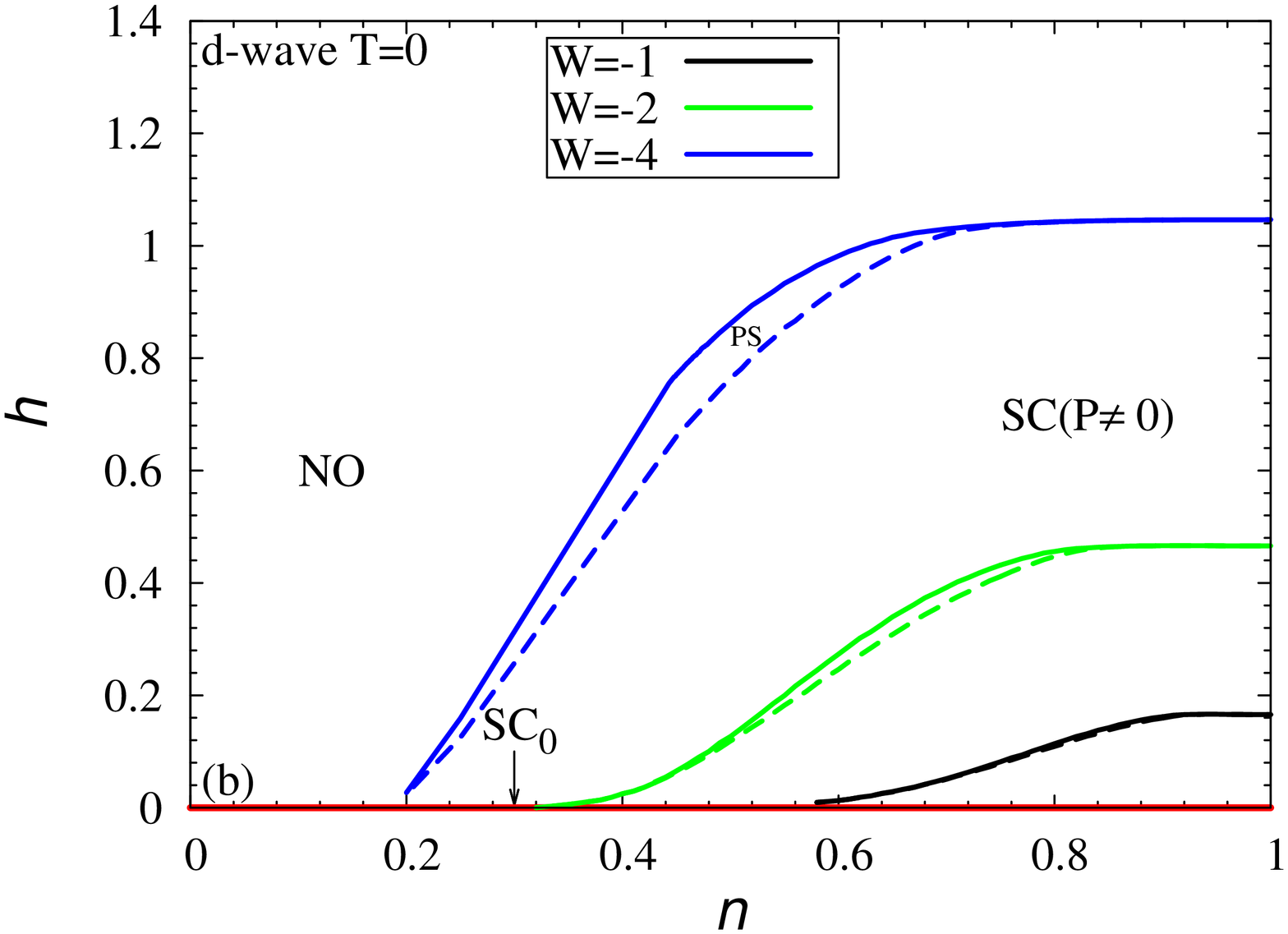}
\caption{\label{diag_d-wave_T0} Critical magnetic field vs. the chemical potential (a) and the electron concentration (b) for the first order SC$(P\neq 0)$-NO transition, at $T=0$; three different values of the attractive interaction.}
\label{fig2}
\end{figure}

Now, let us consider the ground state phase diagrams for the d-wave pairing symmetry case, in the weak coupling regime. Both the fixed chemical potential and the fixed electron concentration case have been analyzed. 

In the s-wave case and in the weak coupling regime, the superconducting
phase with $P=0$ (SC$_0$) is stable at $T=0$. However, the analysis of the
d-wave order parameter behavior, the density of states and the momentum
distributions characteristics \cite{kujawa3} indicate the possibility of the occurrence of stable SC$(P\neq 0)$
phase at $T=0$, even in the weak coupling regime, as opposed to the s-wave
pairing symmetry case in 2D \cite{Kujawa5}. As
shown above, for infinitesimally low value of the magnetic field, the 
state is stable. 
The SC$(P\neq 0)$ phase is the superfluid state of coexisting Cooper
pairs and excess fermions, with the latter responsible for finite
polarization (magnetization) and the gapless excitations characteristic for this
state. 

At higher values of the Zeeman magnetic field, SC$(P\neq 0)$ is destroyed by the
paramagnetic effect or by population imbalance. Then, there is the first order
phase transition from the polarized superconducting phase to the polarized
normal state. The first order transition is manifested by the presence of the
phase separation (PS) region in the phase diagrams at fixed $n$ (see: Fig.
\ref{diag_d-wave_T0}(b)). The phase separation occurs between SC$(P\neq 0)$ with the
number of particles $n_s$ and NO with the number of particles $n_n$. It is worth
mentioning that the d-wave superfluidity is stable around the half-filled
band in the weak coupling limit and its range of occurrence widens with
increasing attractive interaction.  

\begin{figure}[t!]
\includegraphics[width=0.3\textwidth,angle=270]{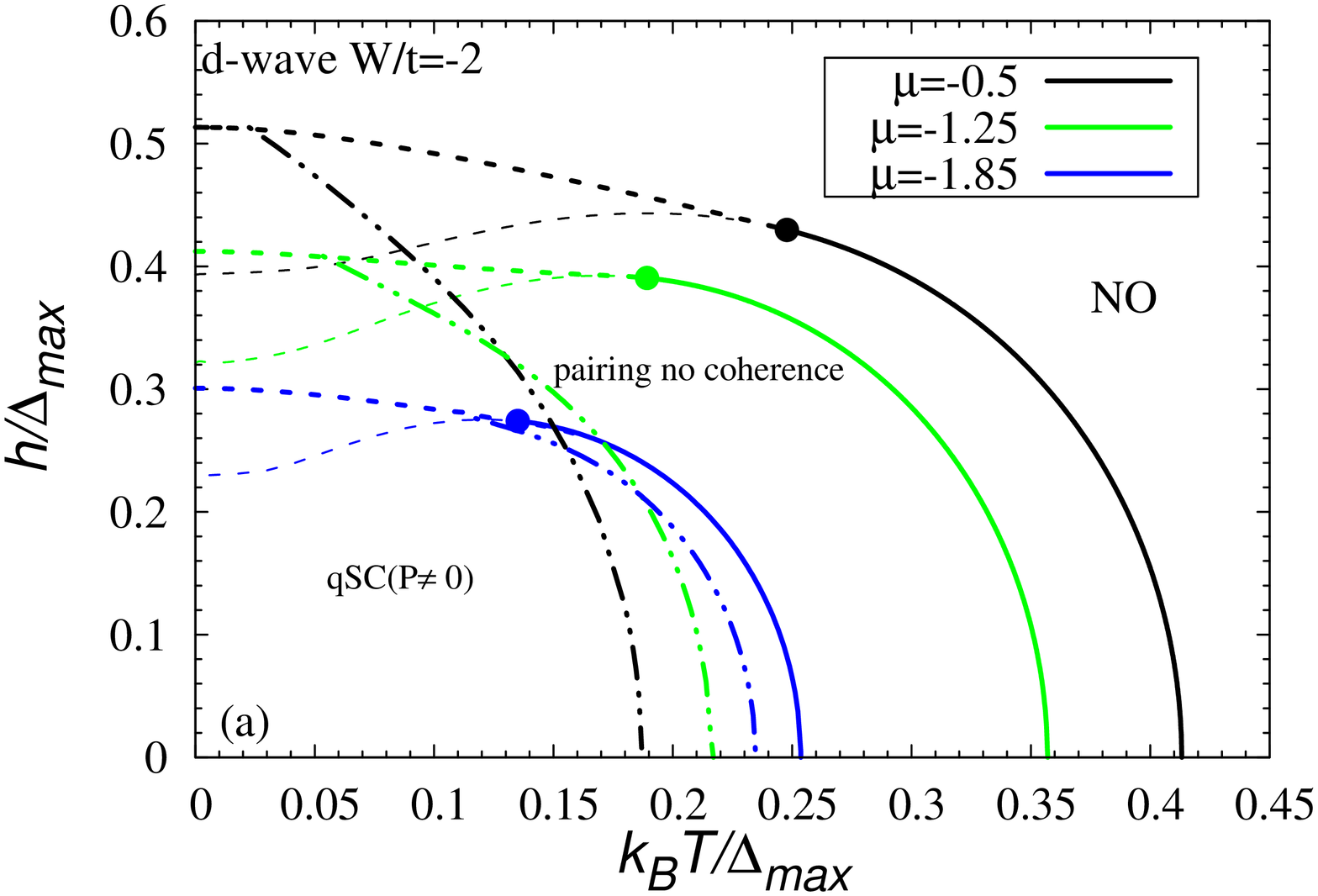}
\includegraphics[width=0.32\textwidth,angle=270]{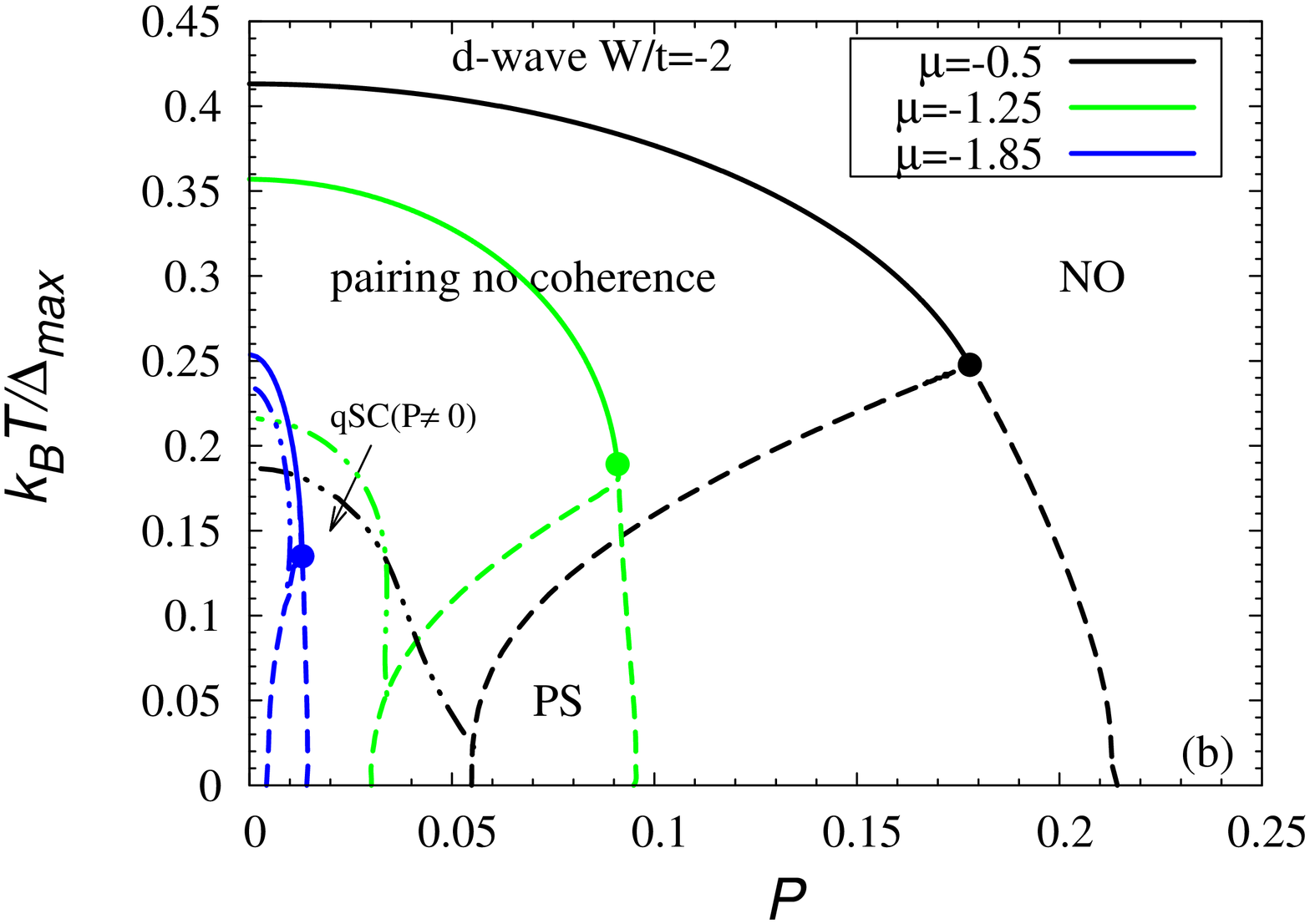}
\caption{\label{diag_T_d-wave} Temperature vs. magnetic field (a) and polarization (b) phase diagrams for $W=-2$, three values of $\mu$; SC$_{0}$ -- non-polarized superconducting state ($P=0$), SC$(P\neq 0)$ -- 2D superconductor in the presence of the polarization, NO -- partially-polarized normal state. The thick solid line is the second order phase transition line from pairing without coherence region to NO. The thin dashed line is an extension of the $2^{nd}$ order transition line (metastable solutions). The thick dashed-double dotted line is the Kosterlitz-Thouless transition line. The thick dotted line denotes the first order phase transition to NO. $\Delta_{0}=4\Delta_{\eta}$ denotes the gap at $T=0$ and $h=0$.}
\end{figure}

Let us discuss the finite temperature phase diagrams.

One of the well-known results concerning the
influence of the Zeeman magnetic field on superconductivity is the existence of the \emph{so-called} Clogston limit \cite{clogston}. In the weak coupling regime, for the s-wave pairing symmetry case, at $T=0$, the
superconductivity is destroyed through the paramagnetic effect and the first
order phase transition to the normal state at a universal value of the critical
magnetic field $h_{c}=\Delta_0/\sqrt{2}\approx 0.707\Delta_0$, where $\Delta_0$
is, the gap at $T=0$ and $h=0$. In turn, this universal value of the
magnetic field in which the superconducting state is destroyed in the ground
state, for the d-wave pairing symmetry case at half-filing is: $h_{c}^{d-wave}=0.56\Delta_{max}$ \cite{Yang}, where
$\Delta_{max}=4\Delta_{\eta}$ at $T=0$ and $h=0$. 

Fig. \ref{diag_T_d-wave} shows the temperature vs. magnetic field ($T-h$) and
polarization ($T-P$) phase diagrams for $W=-2$ and three values of the chemical
potential. These fixed values of $\mu$ correspond to lower values of $n$ than
$n=1$, therefore the d-wave Clogston limit is not reached. However, our results
for $\mu=0$ ($n=1$) agree with the ones from the paper \cite{Yang}, i.e. indeed
$h_{c}^{d-wave}=0.56\Delta_{max}$ for this case.

We take into account the phase fluctuations in $d=2$ within the KT scenario, as
in the s-wave pairing symmetry case \cite{ACichy}. In such way, we can estimate the
phase coherence temperatures, in addition to the mean field (MF) temperatures. The solid
lines (2$^{nd}$ order transition lines) and the PS region are obtained within
the mean field approximation (MFA). The curves below the first order phase transition lines on
the phase diagrams (the thin dotted lines) are the extension of the
2$^{nd}$ order transition lines below tricritical points. The thick dash-double
dotted lines denote the KT transition. The system is a quasi superconductor
(qSC) below $T_c^{KT}$. Between $T_c^{KT}$ and $T_c^{HF}$ pairs still exist, but
without a long-range phase coherence (the pseudogap behavior). The KT
temperatures are much smaller than $T_c^{HF}$. This can be seen particularly
clearly for fixed $\mu=-0.5$ and $W=-2$ case -- the difference between $T_c^{KT}$ and $T_c^{HF}$ amounts to nearly
50\%. However, this difference decreases with decreasing $\mu$ (decreasing $n$)
and decreasing attraction (in the weak coupling limit). The second order phase transition takes place
from the pairing without coherence region to the normal state at sufficiently
low values of the magnetic field. With increasing $h$, the character of the
transition between the pairing without coherence region and the normal state
changes from the second to the first order, which is manifested by the existence
of the MF tricritical point on the phase diagrams. The topology of the ($T-h$) diagrams is the
same as in the s-wave case \cite{Kujawa}.

However, qualitative differences between the s-wave and d-wave pairing
symmetries are clearly visible in $(P-T)$ phase diagrams. At $T\geq 0$ and
$P=0$, there is the unpolarized superconducting phase, both for the s-wave and
the d-wave pairing symmetry case. At $T=0$, there is only the PS region, for the
whole range of polarizations, i.e. $P>0$, for the isotropic order parameter
case. In turn, in the d-wave pairing symmetry case, there is the spin-polarized
superconducting phase at $T=0$, up to some critical value of the
polarization, for which the first order phase transition to the normal state
takes place. In the PS region, not only the polarizations, but also the particle
densities in SC and NO are different. At $T=0$ and for the s-wave pairing
symmetry, this separation region is between the SC$_0$ phase and the normal state, while in the d-wave pairing symmetry it is between SC$(P\neq 0)$ and NO. At $T \neq 0$, $\Delta \neq 0$ and $P\neq 0$, the system is also
in the polarized qSC phase (i.e. homogeneous superconductivity in the presence
of the spin polarization) in the s-wave pairing symmetry case up to $T_c^{KT}$.
The KT phase is restricted to the weak coupling region and low values of $P$, as
in the d-wave pairing symmetry case. Increasing polarization favors the phase of
incoherent pairs. As shown in Fig. \ref{diag_T_d-wave}(b), the range of
occurrence of qSC in the presence of $P$ widens in the weak coupling regime with
increasing $\mu$ (increasing n). In the s-wave pairing symmetry case, one can
distinguish the gapless region at sufficiently high values of the magnetic
field and temperature. As mentioned before, the d-wave pairing symmetry at
$h=0$ is gapless in four nodal points on the Fermi surface. Therefore, the natural
consequence of this is the occurrence of the gapless region also for
infinitesimally low values of $h$, even at $T=0$.     

\section{Conclusions}
We have investigated the influence of a Zeeman magnetic field on the superfluid characteristics of the EHM, within the mean field approximation. We have analyzed the pure d-wave pairing symmetry case. 
At $T=0$, in the presence of the magnetic field, the ground state is the spatially homogeneous spin-polarized superfluid state, which has a gapless spectrum for the majority spin species, for a weak attraction, as opposed to the s-wave pairing symmetry case in 2D. With increasing $h$, the first order phase transition takes place to the NO state.
We have also extended our analysis to finite temperatures in $d=2$ by invoking the KT scenario. 
At finite temperatures, in the weak coupling regime and for fixed $\mu$, the following states have been found in the 2D system: at $h=0$ -- the SC$_0$ phase; at $T=0$, $h\neq 0$ -- polarized superfluid state with a gapless spectrum for the majority spin species; at $T>0$ -- qSC$(P\neq 0)$ (below the Kosterlitz-Thouless temperature); region of pairs without coherence (below the Hartree temperature); the PS region and NO. PS terminates at MF TCP, in ($T-P$) phase diagrams. 
\vspace{-0.7cm}
\begin{acknowledgments}
\vspace{-0.5cm}
I would like to thank R. Micnas for guidance and many valuable discussions.
I acknowledge financial support under grant No. N N202 030540 (MSHE -- Poland).
\end{acknowledgments}

\vspace*{-0.3cm}

\end{document}